\documentstyle[aaspp4]{article}
\begin{document}
 
\lefthead{Elmegreen, Wilcots, \& Pisano}
\righthead{Pattern Speed in NGC 925}
\slugcomment{accepted by ApJ Letters}
 
\title{HI Observations of the Spiral Arm Pattern Speed in the Late-Type
Barred Galaxy NGC 925}
 
\author{
Bruce G. Elmegreen\altaffilmark{1}
Eric Wilcots\altaffilmark{2}, D.J. Pisano\altaffilmark{2}
}
 
\altaffiltext{1}{
IBM Research Division, T.J. Watson Research Center,
P.O. Box 218, Yorktown Heights, NY 10598, USA, bge@watson.ibm.com}
\altaffiltext{2}{University of Wisconsin, Astronomy Department}

\begin{abstract}
HI observations with the VLA of the late-type barred galaxy NGC 925
show clear streaming motions along four spiral arms close to the minor
axis.  A transition from inward streaming to outward streaming is
found, presumably marking the corotation resonance.  The realm of
inward streaming extends to very large radii, corresponding to
$1.3R_{25}$ and 1.7 bar radii deprojected.  This implies that most of
the visible disk is inside corotation, although the HI disk extends
beyond, at least to $\sim2R_{25}$.  As a result, either the bar and
the spirals are rotating separately or the bar rotates very slowly
compared to early-type bars.  There is apparently no inner Lindblad
resonance because of the slowly rising inner rotation curve, and there
may be no outer Lindblad resonance if the disk ends first.
 
\end{abstract}
 
Subject headings: galaxies: individual (NGC 925) --- galaxies: ISM --- 
galaxies: kinematics and dynamics --- galaxies: spiral --- radio
lines: galaxies

\section{Introduction}
NGC 925 is a late-type barred galaxy with deVaucouleurs class SXS7
(=SABd; de Vaucouleurs, et al. 1991 = RC3),
large optical diameter, $D_{25}=10.47'$ (RC3), small distance, 9.3
Mpc (Silberman et al. 1997), and intermediate
inclination, $i=55.8^\circ$ (RC3). It has a bright optical bar
of length $4'$, and two bright, patchy spiral arms
beginning at the ends of the bar and extending out to $\sim3'$ (deprojected)
along the minor axis where they seem to end. It also has HI spiral
arms coincident with the optical arms, but extending to about
twice the radius. The fortuitous positioning of the
optical and radio arms on the minor axis half way out in the disk,
combined with the large size and favorable inclination, makes NGC 925
a good candidate for the observation of radial
streaming motions along the arms, and for the possible determination
of corotation where the radial streaming motions change sign.
 
NGC 925 is important for a study of pattern speeds because it has a
late Hubble type and most determinations of pattern speeds are
for earlier types (see review in Elmegreen 1996).  Early and late-type
bars have different lengths relative to the galaxy sizes and the
rising parts of the rotation curves, and they have different
luminosity profiles (Elmegreen \& Elmegreen 1985;
see review in D.  Elmegreen
1996).  These differences led to the suggestion that early and late
type barred galaxies have different pattern speeds, with corotation
near the end of the bar in early types and further out in late types
(Elmegreen \& Elmegreen 1985).  Computer simulations have obtained
this result too (Combes \& Elmegreen 1993; Noguchi 1996).
 
The distinction between early and late type galaxies in this sense
appears at a Hubble type of about SBbc (D. Elmegreen 1996).
Sempere, Combes, \& Casoli (1995) have in fact found a slowly rotating
pattern for a barred galaxy of this transition
type, NGC 7479, where corotation
is at the edge of the optical disk, at $\sim3$ bar lengths.
NGC 4321 (SABbc) also has spiral arm
corotation well beyond the bar, at $\sim 1.8$ bar lengths (Garcia-Burillo,
Sempere, \& Combes 1994; Sempere et al. 1995). However,
corotation for galaxies as late in type as NGC 925 (SABd) have not
previously been
observed.
 
NGC 925 was observed with the VLA {\footnote{The VLA is part of the
National Radio Astronomy Observatory which is operated by the Associated
Universities Inc., under a cooperative agreement with the National Science
Foundation}}
C configuration in April 1996.  A
detailed discussion of the observations and the overall HI
distribution are in Pisano, Wilcots \& Elmegreen (1997).  Previous HI
observations were reported by Gottesman (1980) and Wevers et al.
(1986).  In what follows, we discuss the streaming motions in the arms
and the implications for the pattern speed of the spiral.
 
\section{Streaming Motions and Corotation}
 
Figure 1 shows the HI column density map as a gray scale and the HI
velocities as contours.  The receding side is west and north is up.
For trailing spirals, this implies that the near side of the galaxy is
in the north.  The positions of various features on the map are
identified by arrows.   Streaming motions are clearly present in the
HI arms.  This is shown in the figure by the local curvature of the
velocity contours inside the arms near the minor axis.
 
The feature labelled $A-B-C$ on the HI map corresponds to the main
optical arm in the south,
which is on the far side of the galaxy.  The velocity contours bend in
a backwards-C pattern, indicating that the velocities near the HI arm
shift toward smaller values.  Small velocities on the far side
minor axis correspond to radial inward motion.  This shift persists
all the way to point $C$ on this arm, and then disappears, presumably
because $C$ is close to the major axis where the line-of-sight
velocity is dominated by rotation.  The magnitude of the line-of-sight
velocity shift on the minor axis is about 10 km s$^{-1}$.
 
The nature of the velocity shift in the southern HI arm is
interesting.
For a small length along the arm, from $A$ to $B$, there is an abrupt
shift toward negative line-of-sight velocity at the inner edge of the
HI concentration, indicative of a shock front that slows down the
outward-moving interarm gas at smaller radii and makes it stream
inwards inside the arm.  Such shock fronts
are rarely observed directly like this.
For a longer length along the HI arm, from $A$ to $C$,
there
is another abrupt shift in velocity at the outer edge of the arm,
indicative of a rarefaction front that returns the gas to its high
outward velocity in the next interarm region.  These sharp features
may be present elsewhere in the spirals, but their observation at this
location is most favorable because the telescope beam (lower left
corner of the map) has its smallest dimension perpendicular to the
arms here.
 
The general pattern of streaming motions in the southern HI arm of NGC
925 is a signature of gas flow inside corotation.  The detailed
features are not the same as those predicted by analytical studies
(Roberts 1969).  The analytical studies for a one-component,
non-self-gravitating fluid in a weak spiral potential have only one
abrupt velocity transition, at the inner edge of the shock, and then a
gradual acceleration of the gas toward higher radial velocities as it
moves through the arm to the interarm.  The same smooth expansion
pattern occurs for cloudy flows (Hausman \& Robert 1984)
which differ primarily
by not having a sharp shock on the inner edge.  Neither of these flows
would give a C-shaped velocity pattern on the minor axis.  The gas
near point $A$ seems to have an equally sharp transition leaving the
HI arm as it does entering the arm, and, where the resolution is poor
and sharp velocity jumps cannot be seen, the velocity contours are
still equally curved on both sides of the arm.  This means that the
gas does not accelerate much to higher radial speeds while it is
inside the HI arm, but maintains about the same speed in the arm and
then accelerates rather abruptly as it leaves.  The same conclusion
could be obtained from the nearly uniform profile of HI column density
across the arm, since the column density is inversely proportional to
the perpendicular speed.  Thus the gas has a broad and nearly uniform
maximum in column density in the HI arm, rather than a single peak at
the shock front and a gradual decrease toward larger radii.  This
uniformity of the arm compression could be the result of self-gravity,
as suggested by Lubow, Cowie, \& Balbus
(1986).
 
The same pattern of velocity shifts, including beam-smeared shock and
rarefaction fronts, is present, although not as prominent, in the northern
arm, from $D$ to $E$,
with a positive line-of-sight velocity shift in this case.  Because
this arm is on the near side of the galaxy, such positive velocity
perturbations indicate inward streaming motions here too.  Thus
section $D$ to $E$ is also inside corotation.
 
The HI column density also shows spiral arms outside these two main
optical arms, in the south from the eastern major axis to just beyond
the minor axis, and in the north from the western major axis nearly
all the way around to the eastern major axis.  There is
streaming motion there too.
 
The southern outer arm is indicated by arrows $F-G-H$, although it
begins slightly to the east of $F$.  Beyond $G$, it is not visible in
the figure as an HI density peak, but it still shows up in the
streaming motions.  Beyond $H$, we lose information about this arm
because the HI is too faint.  The velocity contours in the figure
indicate that the streaming motions are radially inward at $F$ but
radially outward from $G$ to $H$ and beyond.  This is the signature of
a transition from inside corotation to outside corotation in
conventional models, in which case corotation would be near the
southern minor axis between $F$ and $G$.
 
The northern outer arm is indicated by $I-J-K-L$.  The streaming
motions are positive on the line of sight from $I$ to $J$ and negative
from $K$ to $L$, so corotation could be between $J$ and $K$ in the
north.
 
Considering the galaxy inclination of $55.8^\circ$, the galactocentric
distance of a point on the minor axis is $1/\cos 55.8=1.78$ times the
projected distance.  The projected distances of points $G$ and $K$
from the galaxy center at $\delta=33^\circ21.4'$ are $3.0'$ and
$4.0'$, which correspond to 14.4 and 19.3 kpc deprojected for the
assumed galaxy distance
of 9.3 Mpc.  The radius at 25 mag arcsec$^{-2}$ is $R_{25}=14.2$ kpc,
and the deprojected 
optical bar half-length is 5.4 kpc, where the bar ends and the
outer spiral system begins.  If these points $G$ and $K$
are near corotation,
then corotation is at an average
$R_{CR}\sim17$
kpc, which is $\sim3$ bar lengths and $\sim1.2R_{25}$.

The rotation curve for NGC 925 was given in 
Pisano, Wilcots \& Elmegreen (1997). For the deprojected corotation distance
of $6.2'=374''=16.8$ kpc, the rotation speed is $\sim130$ km s$^{-1}$, which 
gives the spiral arms a pattern
speed of $7.7$ km s$^{-1}$ kpc $^{-1}$. 
There is apparently no ILR because of the slowly rising
inner rotation curve, and the OLR is too far out in the disk to observe. 
If the rotation curve remains flat beyond $6.2'$,
then the outer Lindblad resonance is at a distance of 
1.7 times the corotation distance, which 
would correspond to $10.5'$, or 28 kpc.
The apparent lack of an ILR requires confirmation from higher resolution
gas observations, because sometimes CO in the inner disk reveals
a steep rise that was not evident in HI (e.g., 
Sempere \& Garcia-Burillo 1997).
 
\section {Discussion}
 
Stellar orbit theory suggests that corotation of the bar in a barred
galaxy cannot be inside the bar (Contopoulos 1981). As a result,
it was generally assumed that corotation was actually near the end
of the bar.  Now it seems this is only approximately true for early type
galaxies since $R_{CR}\sim1.2R_{bar}$ on average (Elmegreen 1996),
placing corotation in the spiral region. It may not
even be close for late type galaxies, if NGC 925 is typical.  The
implication of our observations 
of NGC 925 
is that 
either the bar and spiral rotate separately (e.g. Sellwood \& Sparke
1988),
or the bar+spiral pattern rotates very slowly compared to early-type bars.

\figcaption{N(HI) in gray and velocities as contours for NGC 925.
Grayscale is from -0.005 to 2.061 KJy Beam$^{-1}$ m s$^{-1}$, and
contours are from 450 to 650 km s$^{-1}$ in steps of
10 km s$^{-1}$.  Labelled arrows indicate ranges of positions for spiral
arm streaming motions, as discussed in the text.}
 

\begin{references}
\reference{} Combes, F., \& Elmegreen, B.G. 1993, A\&A, 271, 391
\reference{} Contopoulos, G. 1980, A\&A, 81, 198
\reference{} de Vaucouleurs, G., de Vaucouleurs, A., Corwin, H.G., Jr.,
Buta, R., Paturel, G., \& Fouque, P. 1991, Third Reference Catalog
of Bright Galaxies, New York: Springer-Verlag
\reference{} Elmegreen, B.G., \& Elmegreen, D.M. 1985, ApJ, 288, 438
\reference{} Elmegreen, B.G. 1996, in Barred Galaxies, ed. R. Buta,
D. Crocker, \& B. Elmegreen, San Francisco: ASP, p. 197
\reference{} Elmegreen, D.M. 1996, in Barred Galaxies, ed. R. Buta,
D. Crocker, \& B. Elmegreen, San Francisco: ASP, p. 23
\reference{} Garcia-Burillo, S., Sempere, M.J., \& Combes, F. 1994,
A\&A, 287, 419
\reference{} Hausman, M.A., \& Roberts, W.W., Jr. 1984, ApJ, 282, 106
\reference{} Lubow, S.H., Cowie, L.L., \& Balbus, S.A. 1986, ApJ,
309, 496
\reference{} Noguchi, M. 1996, ApJ, 469, 605
\reference{} Pisano, D.J., Wilcots, E.M., \& Elmegreen, B.G. 1997, AJ,
submitted
\reference{} Sellwood, J.A., \& Sparke, L.S. 1988, MNRAS, 231, 25p
\reference{} Sempere, M.J., Combes, F., \& Casoli, F.
1995, A\&A, 299, 371
\reference{} Sempere, M.J., Garcia-Burillo, S., Combes, F.,
\& Knapen, J.H. 1995, A\&A, 296, 45
\reference{} Sempere, M.J., \& Garcia-Burillo, S.
1997, A\&A, 325, 769
\reference{} Silberman, N.A., {\it{et al.}} 1996, ApJ, 470, 1.
\end{references}
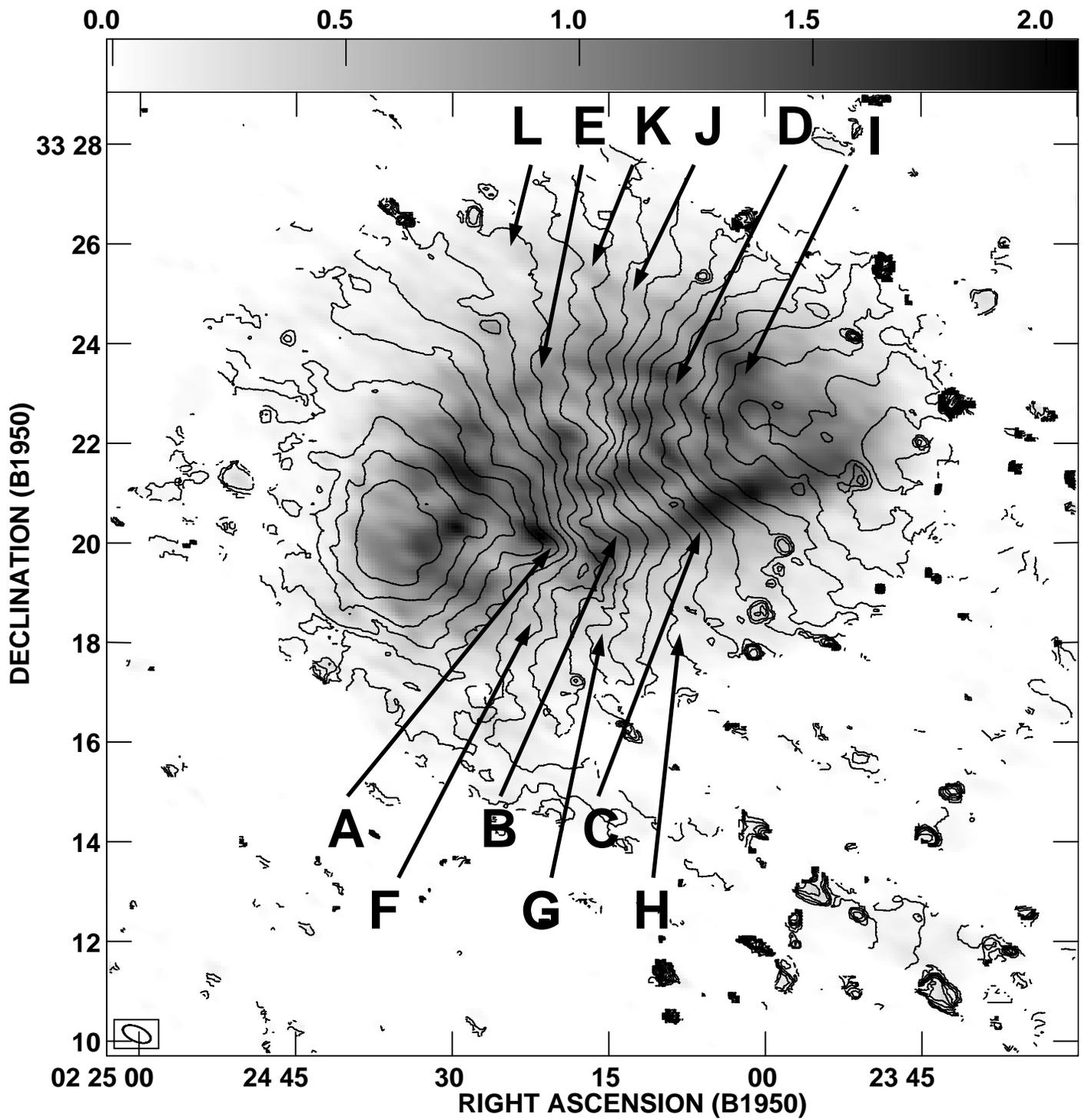
\end{document}